# On the implication of Bell's probability distribution and proposed experiments of quantum measurement

Hai-Long Zhao

Jiuquan satellite launch center, Jiuquan, 732750, China

**Abstract:** In the derivation of Bell's inequalities, probability distribution is supposed to be a function of only hidden variable. We point out that the true implication of the probability distribution of Bell's correlation function is the distribution of the joint measurement outcomes on the two sides. So it is a function of both hidden variable and settings. In this case, Bell's inequalities fail. Our further analysis shows that Bell's locality holds neither for dependent events nor for independent events. We think that the measurements of EPR pairs are dependent events, thus violation of Bell's inequalities cannot rule out the existence of local hidden variable. In order to explain the results of EPR-type experiments, we suppose that polarization entangled photon pair can be composed of two circularly or linearly polarized photons with correlated hidden variables, and a couple of experiments of quantum measurement are proposed. The first uses delayed measurement on one photon of the EPR pair to demonstrate directly whether measurement on the other could have any non-local influence on it. Then several experiments are suggested to reveal the components of polarization entangled photon pair. The last one uses successive polarization measurements on a pair of EPR photons to show that two photons with a same quantum state will behave in the same way under the same measuring condition.

## 1. Introduction

Quantum theory gives only probabilistic predictions for individual events based on the probabilistic interpretation of wave function, which leads to the suspicion of the incompleteness of quantum mechanics and the puzzle of the non-locality of the measurement of EPR pairs [1]. Indeed, if hidden variable theory is not introduced into quantum measurement, we can hardly understand the distant correlation of EPR pairs, e.g. quantum teleportation and quantum swapping [2,3]. Bell pointed out that any theory that is based on the joint assumptions of locality and realism conflicts with the quantum mechanical expectation [4]. Since then, various local and non-local hidden variable models against Bell's inequalities have been proposed (see, e.g. [5-10]), among which the most attractive one is the time-related and setting-dependent model suggested by Hess and Philipp [10], but was criticized by Gill et al. and Myrvold for being non-local [11,12]. As a matter of fact, there is an assumption of probability distribution in the derivation of Bell's inequalities. Bell supposed that it is a function of hidden variable and irrelevant to measuring condition. However, the validity of this assumption is dubious. As pointed out by many authors that if this assumption does not hold, then Bell's inequalities fail [13-15]. On the other hand, it has been shown that even if non-locality is taken into account, Bell's inequalities may also be violated [16,17]. So we focus on Bell's probability distribution and discuss its validity. We point out that its true implication is the probability distribution of the joint measurement outcomes. Since the measurement outcomes are related to settings, its probability distribution is also related to settings. In this case, Bell's inequalities do not hold. We explore the physical meaning of hidden variable and suggest uncertainty of the spatial distribution of the particle as hidden variable.

In terms of quantum entanglement, the spin (polarization) of a pair of EPR particles is indefinite and dependent on each other. By analyzing existing experiments of polarization entanglement [18-31], we show that polarization entangled Bell states (maximally entangled states) can be formed by circularly or linearly polarized photon pairs with correlated hidden variables. If hidden variable does exist, then the quantum state of one of the EPR pair will not change when measurement is made on the other, and the outcomes of a pair of particles with a same quantum state will be the same under the same circumstance. We propose three types of experiments to test above hypotheses. The experiments are easy to realize for the experimental setups are very simple.

## 2. On Bell's probability distribution and suggested hidden variables

Among local hidden variable theories Bell's inequalities play an important role. Bell regarded that his correlation function was founded on the vital assumption of Einstein that the result of $B$ does not depend on the setting of measuring device $a$, nor $A$ on $b$, then it can be written as [4]

$$P(a,b)=\int A(a,\lambda)B(b,\lambda)\rho(\lambda)d\lambda\,, \tag{1}$$

where $A(a,\lambda)=\pm 1$, $B(b,\lambda)=\pm 1$, $\rho(\lambda)$ is the probability distribution of hidden variable according to Bell. de la Peňa et al. suggested that $\rho$ may depend on measuring condition [13]. Nagasawa further expressed this idea by modified definition of locality [14]. But many people insist on the locality of Eq. (1) and they think that the probability distribution of hidden variable cannot be influenced by measuring process. So the arguments of de la Peňa and Nagasawa are not widely accepted. If $\rho$ really represents the probability distribution of hidden variable, then Eq. (1) seems reasonable. Now we analyze the mathematical implication of $\rho$. Eq. (1) includes four joint probabilities, which are $P_{++}(A=1,B=1)$, $P_{+-}(A=1,B=-1)$, $P_{-+}(A=-1,B=1)$ and $P_{--}(A=-1,B=-1)$, respectively. Then we have $P(a,b)=P_{++}-P_{+-}-P_{-+}+P_{--}$. Since $P(a,b)$ actually implies the joint probabilities of the measurement outcomes of $A$ and $B$, $\rho$ must be the joint probability density function with respect to the results of $A$ and $B$, i.e. $\rho=\rho(A=\pm 1,B=\pm 1)$. As the results of $A$ and $B$ depend on the settings of measuring devices and hidden variables of the pair, we have $\rho=\rho(a,b,\lambda)$. If it does not vary with measuring condition, then it becomes the case considered by Bell. For a pair of EPR particles it's easy to understand that they share a same hidden variable. But there is no prior reason that the probability distribution of measurement outcomes is irrelevant to the settings. Two curves are plotted in Fig. 1 representing the possible probability distributions under different measuring conditions $a$, $b$ and $a'$, $b'$, respectively.

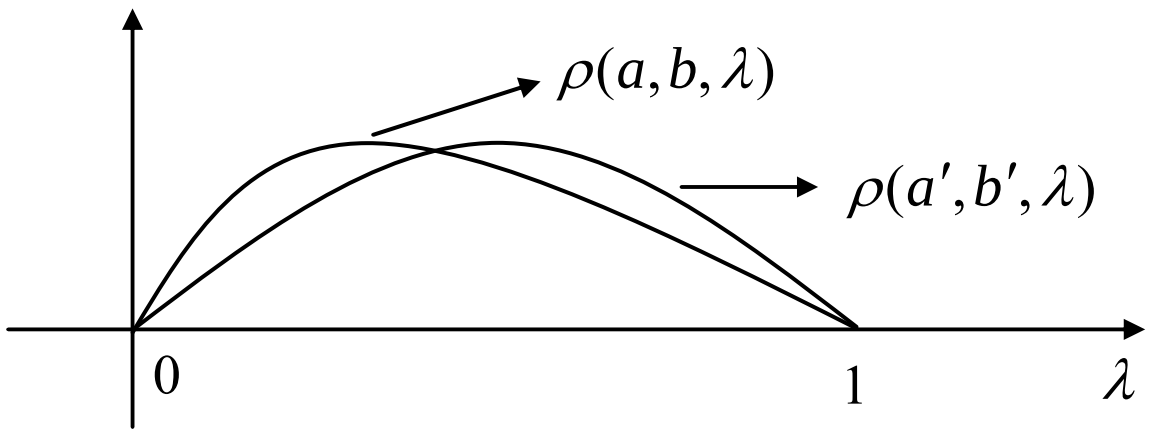

**Fig. 1.** Possible probability distributions under different measuring conditions.

We emphasize that $\rho$ should not be regarded as the probability distribution of hidden variable. Instead, it is the probability distribution of the results $A$ and $B$. Since the joint measurement outcomes are related to $a$, $b$ and $\lambda$, it's natural that the joint probability distribution is a function of $a$, $b$ and $\lambda$. This is the key to understanding Bell's correlation function. It seemed that Bell misunderstood the mathematical implication of the probability distribution.

Now we make further analysis of Bell's correlation function. In the above Bell aimed at the case of a pair of EPR particles. We extend it to the general case where particles $A$ and $B$ have respective hidden variables $\lambda_A$ and $\lambda_B$. As the measurement outcome is related to the local condition and hidden variable, we have $A=A(a,\lambda_A)$ and $B=B(b,\lambda_B)$. In the case that $\lambda_A$ and $\lambda_B$ are mutually independent, we obtain

$$P(a,b)=\iint A(a,\lambda_A)B(b,\lambda_B)\rho(a,\lambda_A)\rho(b,\lambda_B)d\lambda_A d\lambda_B$$

$$=\int A(a,\lambda_A)\rho(a,\lambda_A)d\lambda_A\int B(b,\lambda_B)\rho(b,\lambda_B)d\lambda_B=P_aP_b\,, \tag{2}$$

i.e. joint probability equals the product of individual probabilities, which shows that the two events are independent events. If there exists definite relation between $\lambda_A$ and $\lambda_B$, the two events are dependent events. In this case, joint probability density is not equal to the product of individual probability densities. We can only denote it by $\rho(a,b,\lambda_A,\lambda_B)$. Suppose $\lambda_B = f(a,b,\lambda_A)$, we eliminate integral variable $\lambda_B$ to get

$$P(a,b) = \int A(a,\lambda_A)B(b,\lambda_B)\rho(a,b,\lambda_A,\lambda_B)d\lambda_A = \int A(a,\lambda)B(a,b,\lambda)\rho(a,b,\lambda)d\lambda\,, \tag{3}$$

where $B(a,b,\lambda) = B(b,\lambda_B)$ denotes the result of $B$. For a pair of EPR particles, suppose $\lambda_A = \lambda_B$, we get

$$P(a,b) = \int A(a,\lambda)B(b,\lambda)\rho(a,b,\lambda)d\lambda\,. \tag{4}$$

We see that in this case Eq. (1) should be modified as Eq. (4). Similarly, we have

$$P(a,c) = \int A(a,\lambda)B(c,\lambda)\rho(a,c,\lambda)d\lambda\,, \tag{5}$$

$$P(b,c) = \int A(b,\lambda)B(c,\lambda)\rho(b,c,\lambda)d\lambda\,. \tag{6}$$

With above expressions, Bell's inequalities cannot be obtained. We do not discuss the detailed derivation process.

From above analysis we see that Bell's correlation function holds neither for dependent events nor for independent events. For a pair of EPR particles, their hidden variables may be correlated since they are born from a same particle, so their measurement outcomes are correlated, i.e. the measurements on the two sides are dependent events. Thus violation of Bell's inequalities with EPR-type experiments cannot rule out the existence of local hidden variable.

In the following we discuss the problem of quantum measurement based on the assumption that local hidden variable exists. We first explore the physical meaning of hidden variable. Due to wave-particle duality and uncertainty principle, a microscopic particle may be regarded as a wave packet, which occupies certain volume in space. Hidden variable represents the intrinsic fluctuating state of a particle. So any parameter that can represent the characteristic of spatial distribution of the particle may be used as hidden variable. At present, only the uncertainties of position, momentum and angular momentum et al. may be used to denote this property, so we might as well borrow them to represent hidden variables. Note that the intrinsic quantum fluctuation of the particle is not random, it also obeys certain laws which are unknown to us.

Take spin (polarization) of a particle as an example. In classical theory angular momentum is a vector, whose magnitude and the projections in three directions are all well-defined. In quantum mechanics, the magnitude of angular momentum is well-defined, and we can determine its projection $l_z$ in one direction. But the angular position $\phi$ and the other two projections $l_x$ and $l_y$ are all indefinite. $\phi$ and $l_z$ satisfy the uncertainty relation $\Delta\phi\Delta l_z \geq \hbar/2$. Both $\Delta\phi$ and $\Delta l_z$ indicate the quantum fluctuation of a particle around the projection (measurement) direction, so they may be used as hidden variables. As spin (polarization) is a relativistic quantum effect, it's likely that the corresponding hidden variables are irrelevant to time. We will test this hypothesis in the following experiment.

The hidden variables of spin (polarization) represent the particle's quantum fluctuation of the degree of freedom of spin (polarization) in three-dimensional space, which should be independent of external circumstance. However, the measurement on the particle always projects the spin (polarization) onto a specific direction. The quantum fluctuation of spin (polarization) may be different in different directions, i.e. hidden variable is multi-valued. In this sense, we may also think that hidden variable varies with measuring condition. We now try to explore the measuring process. In classical mechanics and quantum field theory, we have principle of least action. We may introduce this principle into quantum measurement. We define $\Delta\phi\Delta l_z$ as the action for spin (polarization) of a particle in the projection (measurement) direction. When a photon is incident on a polarizer, it has two choices. Consequently, there are two possible collapsed polarization directions and two corresponding actions. We suppose a photon always chooses the direction with a less action. For a linearly polarized photon, its polarization direction may be regarded as the direction with the least action, i.e. in this direction we have $\Delta\phi\Delta l_z = \hbar/2$. Thus when the polarization direction of a photon is parallel to the orientation of a polarizer, it will pass through the polarizer with certainty. Similarly, we define the product of the uncertainties of position and momentum as the action for the motion of center of mass of a photon.

In the general case, when measurement is made on a particle, its quantum state will collapse into another one, and the collapsing process is nonlinear and irreversible. A small change of external circumstance or hidden variable may lead to a different result, i.e. the measurement outcome is sensitive to external circumstance and hidden variable. So the collapse of quantum state is chaotic. From this point of view, the evolutions of microcosm and macrocosm, and even the universe are chaotic in essence.

### 3. Interpretation of EPR-type experiment

The experiment used to test Bell's inequalities with polarization state of photon pairs is shown in Fig. 2. A pair of EPR photons is incident on a pair of polarization analyzers $a$ and $b$, respectively. We denote the transmitted and reflected channels by "+" and "–", respectively. The results for $|\phi^+\rangle$ state in quantum mechanics are [24]

$$P_+(a) = P_-(a) = 1/2\,, \tag{7}$$

$$P_+(b) = P_-(b) = 1/2\,, \tag{8}$$

$$P_{++}(a,b) = P_{--}(a,b) = \frac{1}{2}\cos^2(a-b)\,, \tag{9}$$

$$P_{+-}(a,b) = P_{-+}(a,b) = \frac{1}{2}\sin^2(a-b)\,, \tag{10}$$

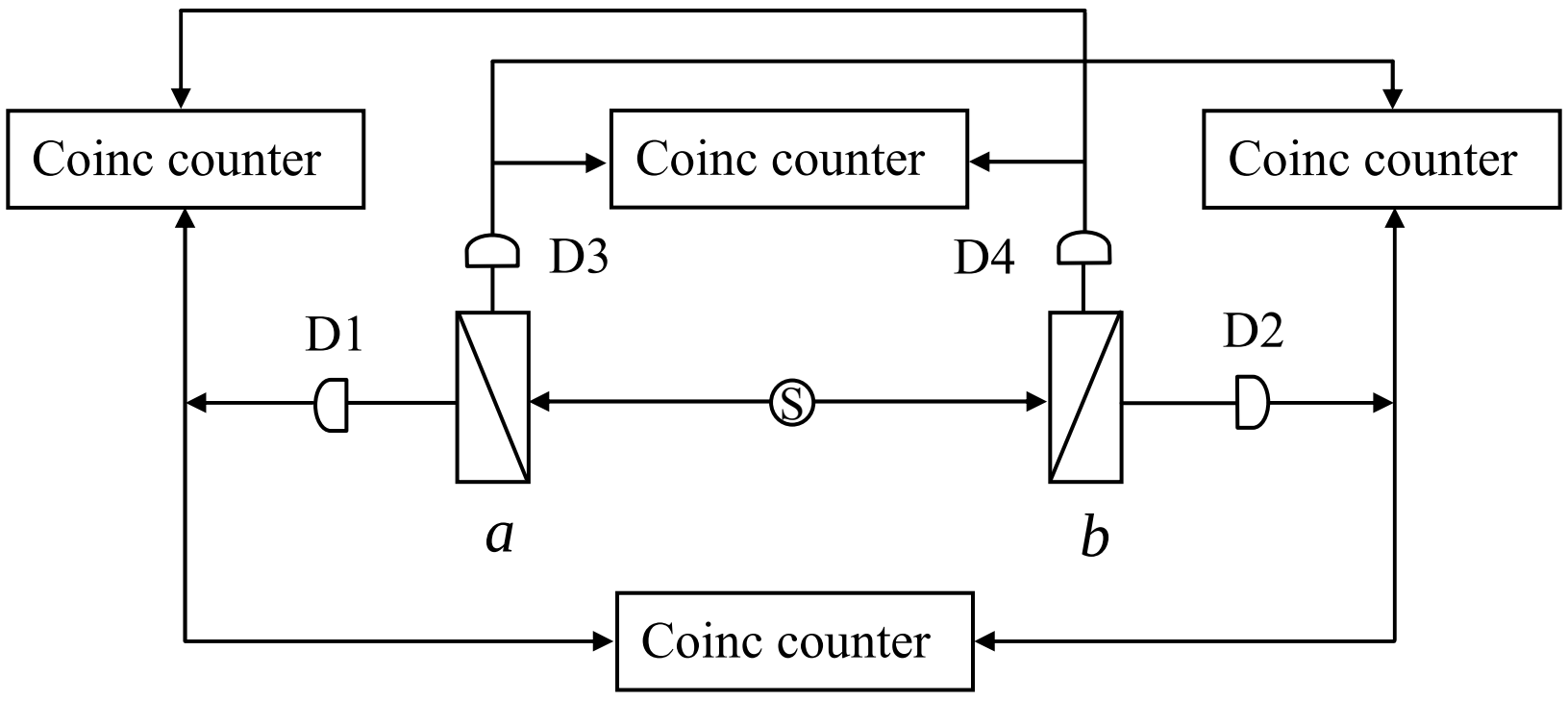

**Fig. 2.** Experimental test of Bell's inequalities.

respectively. In terms of quantum entanglement the polarization of a pair of EPR photons is indefinite. If hidden variable exists, the polarization of each photon should be well-defined. Consider the experiment of photon pairs emitted by the $J=0 \to J=1 \to J=0$ cascade atomic calcium [18,19]. According to classical theory, the two photons are circularly polarized. For the experiment of $J=1 \to J=1 \to J=0$ cascade atomic mercury [20], one photon is linearly polarized and the other circularly polarized. In the case of down-conversion of nonlinear crystal [21-31], the wave packets of two orthogonally polarized photons overlap at crystal or beam splitter. They will form two circularly polarized photons under appropriate conditions. The combination of a half-wave plate and a quarter-wave plate can transform a Bell state into other three Bell states [24]. From these facts, we think that Bell state can be composed of circularly (or circularly/linearly) polarized photon pairs. For the twin photons generated in cascade radiation or down-conversion, their hidden variables may be regarded as correlated, so measurements on the two photons are dependent events. In order to obtain the joint probabilities, we use projective geometry to calculate the conditional probabilities.

We first consider Bell state composed of circularly polarized photon pairs. For a circularly polarized photon, the probabilities of being transmitted and reflected are both 1/2 no matter how we orientate the polarizer. Thus for single probabilities we get the results of Eqs. (7) and (8). For a pair of correlated photons, we may use conditional probability to get

$$P_{++}(a,b) = P_{+}(a)P_{+}(b\,|\,a) = P_{+}(b)P_{+}(a\,|\,b)\,, \tag{11}$$

where $P_{+}(b\,|\,a)$ and $P_{+}(a\,|\,b)$ are conditional probabilities, which can be calculated by projective method. For $|\phi^{+}\rangle$ state we suppose $P_{+}(b\,|\,a) = P_{+}(a\,|\,b) = \cos^2(a-b)$. We may understand above method as follows. If the photon on the left side can pass through the polarizer $a$, then the photon on the right side can certainly pass through a polarizer with the same orientation. If the orientation of the polarizer on the right side is set at $b$, the probability that the photon on the right side can pass through the polarizer is $\cos^2(a-b)$. Then we have $P_{++}(a,b) = \frac{1}{2}\cos^2(a-b)$, which agrees with Eq. (9). Note that only for a pair of circularly polarized photons with maximally correlated or anti-correlated hidden variables ($\lambda_A = \lambda_B$ or $\lambda_A = -\lambda_B$) can we use this projective method. For circularly polarized photon pairs with independent hidden variables, we have $P_{++}(a,b) = P_{+}(a)P_{+}(b) = 1/4$.

As for the Bell state composed of circularly/linearly polarized photon pairs, we suppose the circularly polarized photons are incident on polarizer $a$ and linearly polarized photons on polarizer $b$. We first project $a$ onto $b$. As $P_{+}(a) = 1/2$ and the angle between the orientations of the two polarizers is $a-b$, we use projective geometry to get $P_{++}(a,b) = \frac{1}{2}\cos^2(a-b)$. We then project $b$ onto $a$. Suppose the polarization directions of linearly polarized photons distribute uniformly in space and the angle between the polarization direction of a photon and the orientation of polarizer $b$ is $x$. The probability that a photon can pass through polarizer $b$ is $\cos^2(b-x)$ according to Malus' law, then the joint probability is

$$P_{++}(a,b) = \frac{1}{2\pi}\int_0^{2\pi} \cos^2(b-x)\cos^2(a-b)dx = \frac{1}{2}\cos^2(a-b)\,. \tag{11}$$

If the polarization directions of linearly polarized photons distribute only in two orthogonal directions, we have

$$P_{++}(a,b) = \frac{1}{2}\cos^2 x\cos^2(a-b) + \frac{1}{2}\sin^2 x\cos^2(a-b) = \frac{1}{2}\cos^2(a-b)\,, \tag{12}$$

which also agrees with the result of quantum mechanics. Additionally, if the linearly polarized direction of photons is set at the $\pm 45^\circ$ direction relative to the orientation of the polarizer, the probabilities that the linearly and circularly polarized photons can pass through their respective polarizers are both 1/2. In this case, we also get the same result as that of quantum mechanics using projective method.

In the general case, linearly polarized photon pairs cannot form a Bell state (which we will discuss in detail in the next section). But in special case their joint probability may also agree with the result of quantum mechanics. Suppose a pair of photons has the same polarization direction and the polarization directions of photon pairs distribute in two orthogonal directions with equal probability, and the orientation of polarizer $a$ is in the $x$ (or $y$) direction, while the orientation of polarizer $b$ may vary arbitrarily. When the polarization of a pair of photons is in the $x$ (or $y$) direction, the photon incident on polarizer $a$ can pass through with certainty, and the probability that the photon incident on polarizer $b$ can pass through is $\cos^2(a-b)$ according to Malus' law. When the polarization of a pair of photons is in the $y$ (or $x$) direction, the photon incident on polarizer $a$ cannot pass through. So the joint probability for the photon pairs to pass through the polarizers is $P_{++}(a,b)=\frac{1}{2}\cos^2(a-b)$. In this case, linearly polarized photon pairs can also form a Bell state.

We summarize as follows: (i) circularly polarized photon pairs with correlated hidden variables will form a Bell state; (ii) circularly/linearly polarized photon pairs with correlated hidden variables can form a Bell state under the condition that the polarization directions of linearly polarized photons distribute uniformly in space or in two orthogonal directions, or the linearly polarized direction of photons is set at the $\pm 45^\circ$ direction relative to the orientation of the polarizer; (iii) linearly polarized photon pairs with correlated hidden variables can form a Bell state only when the polarization directions of photon pairs distribute in two orthogonal directions with equal probability and the orientation of one of the polarizers is parallel to one of the polarization directions of the photon pairs.

We have supposed above that the measurement outcome of a photon is determined by the external condition and hidden variable. In fact, it may also be determined by other property of the photon. Consider the Bell state composed of circularly polarized photon pairs. Even if the polarization uncertainties of a pair of photons are the same, their rotation directions may be different. We denote the hidden variables of a pair of photons by $\lambda_A$ and $\lambda_B$, respectively, and the rotation directions of the pair by $d_A$ and $d_B$, respectively. Then the four Bell states can be denoted by the combination of $\lambda$ and $d$. Let's suppose that for $|\phi^+\rangle$ state we have $\lambda_A=\lambda_B$ and $d_A=d_B$ while for $|\psi^-\rangle$ state we have $\lambda_A=-\lambda_B$ and $d_A=d_B$. The coincidence rate of $P_{++}$ for the four Bell states $|\phi^+\rangle$, $|\phi^-\rangle$, $|\psi^+\rangle$ and $|\psi^-\rangle$ are $\frac{1}{2}\cos^2(a-b)$, $\frac{1}{2}\cos^2(a+b)$, $\frac{1}{2}\sin^2(a+b)$ and $\frac{1}{2}\sin^2(a-b)$, respectively [24]. Then we may infer that the rotation direction determines the sign of plus or minus while the hidden variable determines the expression of sine or cosine. So for $|\psi^+\rangle$ state we have $\lambda_A=-\lambda_B$ and $d_A=-d_B$, and for $|\phi^-\rangle$ state we have $\lambda_A=\lambda_B$ and $d_A=-d_B$. As the rotation direction of the photon is a measurable quantity, we do not regard it as a hidden variable.

As for the Bell states composed of circularly/linearly (or linearly) polarized photon pairs, we may use polarization uncertainty and one of the polarization components (e.g. horizontal or vertical polarization) of the pair to denote the four Bell states. For example, $|\phi^+\rangle$ state may be denoted by $\lambda_A=\lambda_B$ and $H_A=H_B$ (or $V_A=V_B$). As for $|\phi^-\rangle$ state, we have $\lambda_A=\lambda_B$ and $H_A=-H_B$ (or $V_A=-V_B$).

We now use above theory to explain the experimental results. The atomic cascade radiation experiments in Refs. [18-20] can be explained by circularly polarized photon pairs or circularly/linearly polarized photon pairs. For the down-conversion of crystal, the wave packets of a pair of orthogonally polarized photons overlap at crystal or beam splitter. If their phases are the same, when they are separated from each other at the output port, they tend to convert into a pair of circularly polarized photons with different rotation directions. As the two photons have anti-correlated hidden variables, the experiment will generate $|\psi^+\rangle$ state. If the two photons obtain phase shift of $\pm\pi/2$ during the propagation process in the crystal due to their different phase velocities, they will form a pair of circularly polarized photons with the same rotation direction. Then the experiment will generate $|\psi^-\rangle$ state. This can explain the experimental results of Refs. [21-27]. As for the Bell state composed of four photons, provided that a pair of orthogonally polarized photons can form a pair of circularly polarized photons with the same rotation direction, Bell state will be obtained. Even if each pair of photons forms a pair of linearly polarized photons with polarization direction at $\pm 45^\circ$, we can turn them into circularly polarized photons by inserting two quarter-wave plates into the optical paths. This can explain the experimental results of Refs. [28-31]. In the meanwhile, As the quantum states of a pair of photons in the same path are the same, even if one photon is lost during the detection process, the coincidence rate will remain unaffected. This type of experiments can increase the detection efficiency of correlated photon pairs.

We have supposed that a pair of linearly polarized photons will be decomposed of a pair of circularly polarized photons with different rotation directions when their wave packets are separated from each other. Certainly, they may also convert into a pair of linearly polarized photons with the polarization direction of $\pm 45^{\circ}$. In this case, if one of the orientations of the polarizers is set at $\pm 45^{\circ}$, Bell state will also be obtained based on our above analysis. As in most of the experiments, one of the polarizer is oriented at $\pm 45^{\circ}$, this possibility cannot be ruled out. In order to test which assumption is correct, we let the orientation of the fixed polarizer deviate from $\pm 45^{\circ}$, e.g. $\pm 20^{\circ}$, while the orientation of the other polarizer may vary arbitrarily. If Bell state can still be obtained in this case, then the former assumption is correct, otherwise the latter is correct. As some experiments have already indicated that Bell state can be obtained when the fixed polarizer is orientated at $0^{\circ}$ or $90^{\circ}$ [25,26], it's likely that the former assumption is correct. In the meanwhile, this experiment provides a method to discriminate between quantum theory and our theory. In the experimental setups of Refs. [21-26], we insert a quarter-wave plate into each optical path. According to quantum theory, the Bell state will remain unaffected. While in our theory, linearly polarized photon pairs will be obtained by inserting quarter-wave plates, and Bell state cannot be obtained in the general case. Then we can decide which theory is correct based on the experimental results.

## 4. Proposed experiments of quantum measurement

### *4.1. Experimental test of the locality of the measurement of EPR pairs*

One of the questions raised by EPR paradox is: if we have measured one particle of the EPR pair, what is the quantum state of the other? For example, suppose $|\phi^{+}\rangle$ state is composed of circularly polarized photon pairs. According to quantum entanglement, when we measure one photon find it linearly polarized, the other will instantaneously collapsed into linear polarization. In terms of hidden variable theory, the other will remain circularly polarized until we analyze it with a polarizer. Does this violate the conservation of angular momentum? If we only consider the system composed of a pair of photons, the angular momentum of the system is certainly not conserved. In the measuring process, a third component—the measuring device is involved. If the measuring device is included, the momentum and angular momentum of the system are still conserved.

In order to discriminate between the two hypotheses, we must seek a material that can exhibit different effects when circularly and linearly polarized photons pass through it respectively. Note that the usual method of inserting a quarter-wave plate into the optical path cannot be used here for the circularly polarized photons in one optical path may have two rotation directions, so we make use of roto-optic effect (or Faraday effect). This is because a linearly polarized photon can be regarded as the combination of left-handed and right-handed circularly polarized components. When it passes through a roto-optic material, the velocities of the two components are different according to Fresnel's roto-optic theory. Then there exists a phase shift between the two components. The polarization plane of the photon will rotate and the polarization quantum state will change. As a circularly polarized photon passes through the roto-optic material, its polarization quantum state will not change since it has only one rotation direction. The experimental setup is shown in Fig. 3, where I and II are a pair of polarizers with the same orientation, and Ro is a roto-optic material which rotates the polarization plane of linearly polarized photon by $\pi/2$. $|\phi^{+}\rangle$ state composed of circularly polarized photon pairs can be generated by down-conversion of nonlinear crystal. When the wave packets of two orthogonally polarized photons overlap at beam splitter or crystal [21,22,24], we may think that $|\psi^{+}\rangle$ state generated in the experiments is composed of circularly polarized photon pairs. Then $|\phi^{+}\rangle$ state can be obtained by inserting a half-wave plate into one of the optical paths. A circularly polarized photon will remain circularly polarized after it passes through a half-wave plate. Thus $|\phi^+\rangle$ state obtained in this way is composed of circularly polarized photon pairs. If we adopt the method of cascade radiation, then the experiments in [18,19] just generate $|\phi^+\rangle$ state. Let the distance between source S and Ro be longer than that between S and polarizer I (L2>L1). Then the leftwards-traveling photon will first be analyzed. Co is an optical path length compensator used to guarantee the simultaneous detection of a pair of photons within the coincidence time window of the counters D1 and D2. If roto-optic material is a Faraday rotator, then the compensator can be used with another same one that is power-off. As a matter of fact, if the optical path length difference between the two sides is appropriately adjusted, the compensator Co may be removed.

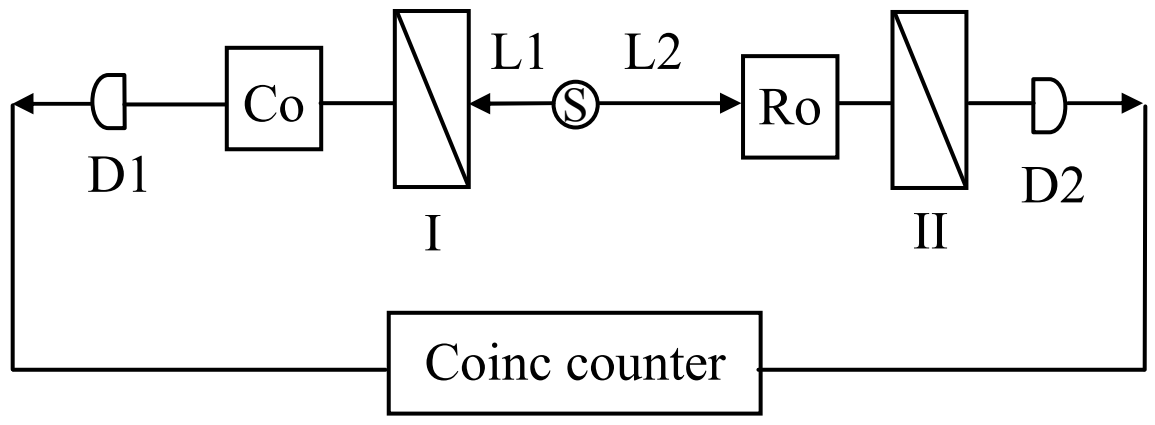

**Fig. 3.** Experimental test of the locality of the measurement of EPR pairs.

We now see the expectations of the two theories. According to quantum entanglement, when the leftwards-traveling photon passes through polarizer I, the polarization direction of the rightwards-traveling photon will instantaneously collapse to the orientation of polarizer I. Its polarization plane is then rotated by $\pi/2$ when it passes through Ro. Thus it will be reflected by polarizer II. If the leftwards-traveling photon is reflected by polarizer I, the coincidence rate is zero whatever the rightwards-traveling photon is transmitted or reflected. So the expected coincidence rate is zero in terms of quantum entanglement. According to hidden variable theory, measurement on one photon does not affect the other. On the other hand, roto-optic material does not change the polarization quantum state of circularly polarized photon. So the coincidence rate will remain unchanged and is always 1/2. If hidden variable varies with time, as suggested by Hess and Philipp [10], the coincidence rate will vary with the position of polarizer II. Similar experiments can be performed for the other three Bell states.

If one does not agree with the assumption of wave packet reduction of EPR pair and supposes roto-optic material does not change the polarization quantum state of EPR pair, he will get the same result as ours. In order to see whether roto-optic material can change the polarization quantum state of EPR pair or not, we make the above experiment with $|\phi^+\rangle$ state composed of circularly and linearly polarized photon pairs. Then a question arises: how to obtain this quantum state? When the wave packets of two orthogonally polarized photons overlap at beam splitter, $|\psi^+\rangle$ state will be generated. Then the two photons are circularly polarized. In the experimental setup of Ref. [23], the rotation direction of one photon is reverted by a reflected mirror, so the experiment will generate $|\psi^-\rangle$ state. Then we can change it into $|\phi^+\rangle$ state with a half-wave plate and a quarter-wave plate. A quarter-wave plate will transform circular polarization into linear polarization, so in this case $|\phi^+\rangle$ state is composed of circularly and linearly polarized photon pairs. Similarly, in the experimental setup of Ref. [24], we let the experiment generate $|\psi^-\rangle$ state by adjusting the birefringent phase shifter. We then use a half-wave plate and a quarter-wave plate to change $|\psi^-\rangle$ state into $|\phi^+\rangle$ state. In this case, $|\phi^+\rangle$ state is composed of circularly and linearly polarized photon pairs. If Ro is inserted into the optical path without quarter-wave plate (the photons in this path are circularly polarized), both theories expect the coincidence rate to be 1/2. However, if Ro is placed into the optical path with quarter-wave plate, the expectations of the two will be different. If roto-optic material does not change the polarization quantum state, the coincidence rate will remain unchanged. According to our theory, the roto-optic material acts as a half-wave plate since it rotates the polarization plane by $\pi/2$, which will transform $|\phi^{+}\rangle$ into $|\psi^{+}\rangle$ state, so we expect the coincidence rate to be $\frac{1}{2}\sin^{2}(a+b)$.

In Wheeler's delayed-choice experiments (e.g. [32-34]), which-way measurements are made with a two-path interferometer which is chosen after a single-photon pulse entered it. The experiments support Bohr's statement that the behavior of a quantum system is determined by the type of measurement, but cannot answer the question as to whether measurement on one particle of EPR pair can affect the other or not. The above experiments can unambiguously answer it and help to understand EPR paradox (GHZ theorem and Hardy theorem as well), which supposes that we can predict with certainty a particle's quantum state by measuring its partner. The above experiments will show that this is not always possible. For example, if we measure photon $A$ with a polarizer and find it to be in $|H\rangle$ state, then photon $B$ may be neither in $|H\rangle$ state nor in $|V\rangle$ state. Instead, it may remain in the superposition state, i.e. circular polarization. Only after measurement with a polarizer can we obtain its definite polarization state ($|H\rangle$ or $|V\rangle$ state), and different measurements will lead to different results. So the hypothesis of EPR paradox is not correct.

### *4.2. Experimental test of the components of polarization entangled photon pairs*

We have supposed above that polarization entangled Bell states can be composed of circularly polarized photon pairs. To test this assumption, we use a pair of linearly polarized photons generated by type-I non-collinear down-conversion. The experimental setup is shown in Fig. 4. Since the two photons are generated from a same photon, their hidden variables should be correlated. Two quarter-wave plates are inserted into the optical paths to convert the linearly polarized photons into circular polarized ones. If the optical axes of the two quarter-wave plates are parallel, the experiment will generate $|\phi^{+}\rangle$ state. If the optical axes are oriented orthogonally, i.e. one is set at $45^{\circ}$, the other at $-45^{\circ}$, the rotation directions of the two circularly polarized photons are opposite, then $|\phi^{-}\rangle$ state will be obtained. Similar experiment can be made with type-II non-collinear down-conversion.

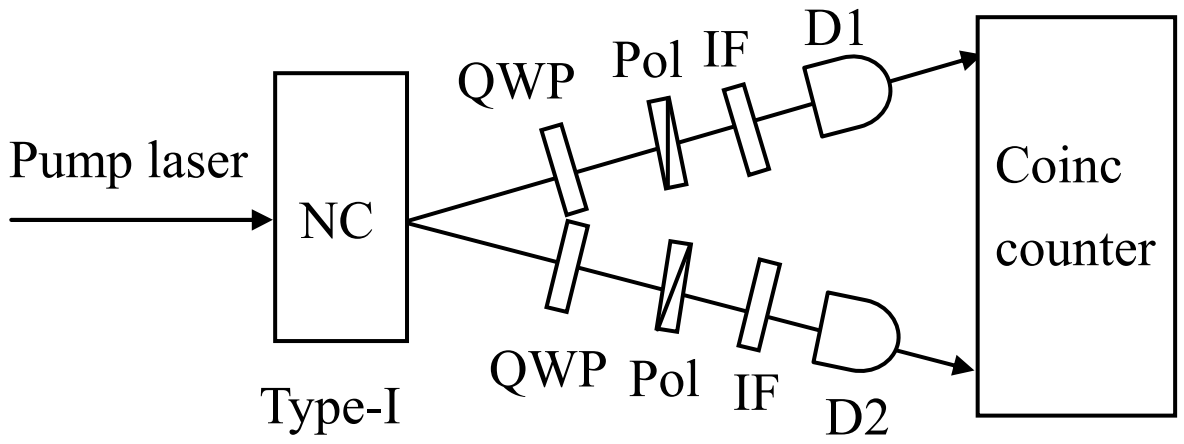

**Fig. 4.** Generation of $|\phi^{\pm}\rangle$ state by type-I non-collinear down-conversion.

For type-II collinear down-conversion, the hidden variables of the two photons may be regarded as maximally anti-correlated. In this case, a polarizing beam splitter (PBS) may be used to separate the two orthogonally polarized photons. Then $|\psi^{\pm}\rangle$ state can be obtained with two quarter-wave plates after the PBS ($|\psi^{+}\rangle$ state will be generated when the optical axes of the two quarter-wave plates are parallel). The experimental setup is shown in Fig. 5.

In order to verify the assumption that circularly and linearly polarized photon pairs can form a Bell state, we remove a quarter-wave plate in the experiment of Fig. 4 or 5, and set the orientation of the polarizer at the $\pm 45^{\circ}$ direction relative to the linearly polarized direction of photons, while the other orientation of the polarizer may vary arbitrarily. In this case, we still obtain $|\phi^{\pm}\rangle$ state in Fig. 4 and $|\psi^{\pm}\rangle$ state in Fig. 5.

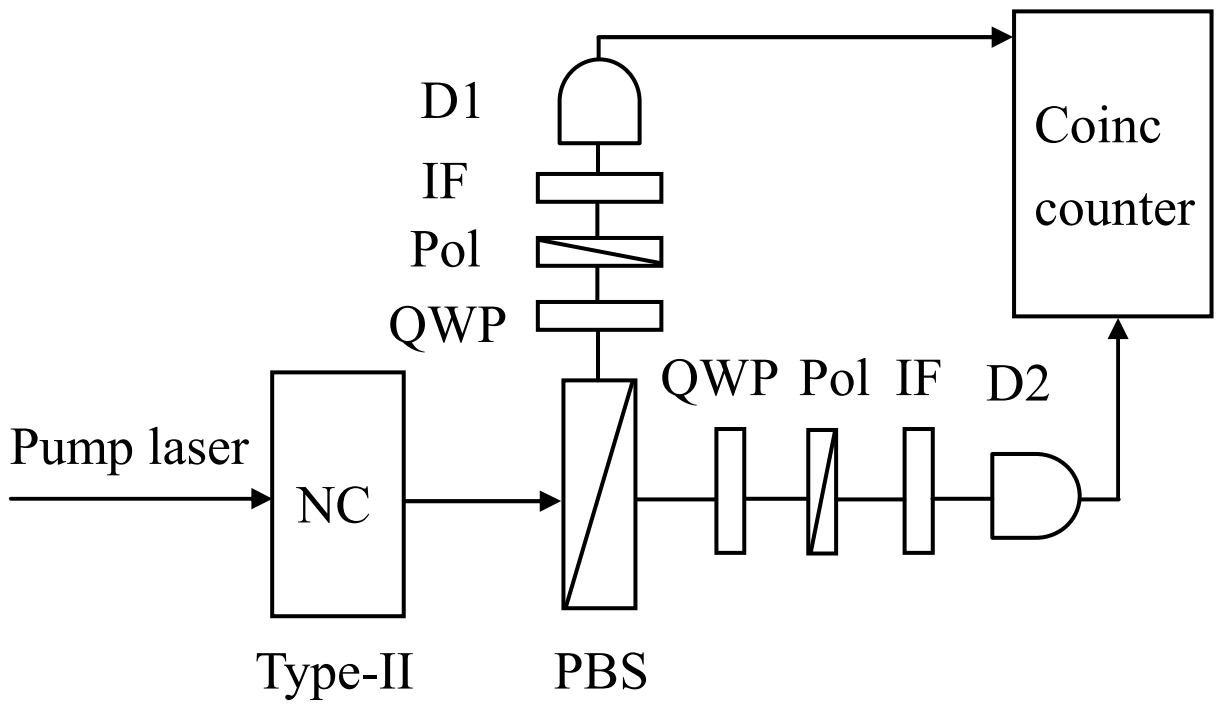

**Fig. 5.** Generation of $|\psi^{\pm}\rangle$ state by type-II collinear down-conversion.

In other down-conversion experiments [21-31], the wave packets of two orthogonally polarized photons overlap at beam splitter or crystal. The above experiments do not overlap the wave packets of photons and the polarizations of photons are definite. If Bell states can be generated in this way, then quantum entanglement will not remain a mystery.

The following experiment uses the overlap of multi-photon wave packets to generate Bell state. The experimental setup is shown in Fig. 6. A beam of linearly polarized laser enters Mach-Zehnder interferometer, which may be continuous-wave or pulsed laser. A half-wave plate is inserted into one of the arms to rotate the polarization plane by $\pi/2$. If we replace the first (or second) BS with a PBS, the half-wave plate can be removed, and the polarization of input laser should be set at $\pm 45^{\circ}$. If the relative phase of the photons in the two arms is correctly chosen, the output will be circularly polarized. On the other hand, since the photons within coherence length are coherent, or indistinguishable, we may think that the polarization hidden variables of a bunch of photons within coherence length are correlated. So these photons will behave in the same way when analyzed by a polarizer, i.e. if one photon is transmitted, then all the photons will be transmitted. In the case that all the photons within the coincidence time window of the detectors are coherent, Bell state will be obtained. Note that the experiment adopts multi-photon wave packets overlap, so similar to the experimental results of the overlap of two biphoton wave packets at beam splitter or crystal [28-31], we expect the experiment will generate $|\phi^{\pm}\rangle$ state. A glass plate may be inserted into the other arm or we can scan one of the mirrors to change the relative phase of the photons in the two arms, and the optical path length difference should be shorter than the coherence length of laser. The key to the experiment is that we must ensure that the polarization quantum states of the photons within the detection time of the detectors are identical, otherwise the behaviors of a bunch of photons will be different. For continuous-wave laser, the coincidence time window of the photon detectors should be shorter than the coherence time of laser. While for pulsed laser, the coherence time of photons should be longer than (or equal to) the duration of pulse, which can be realized by inserting an interference filter in front of each of the detectors. Compared with other beam splitter schemes to obtain Bell states, the experiment is much simpler for it does not use down-conversion of crystal.

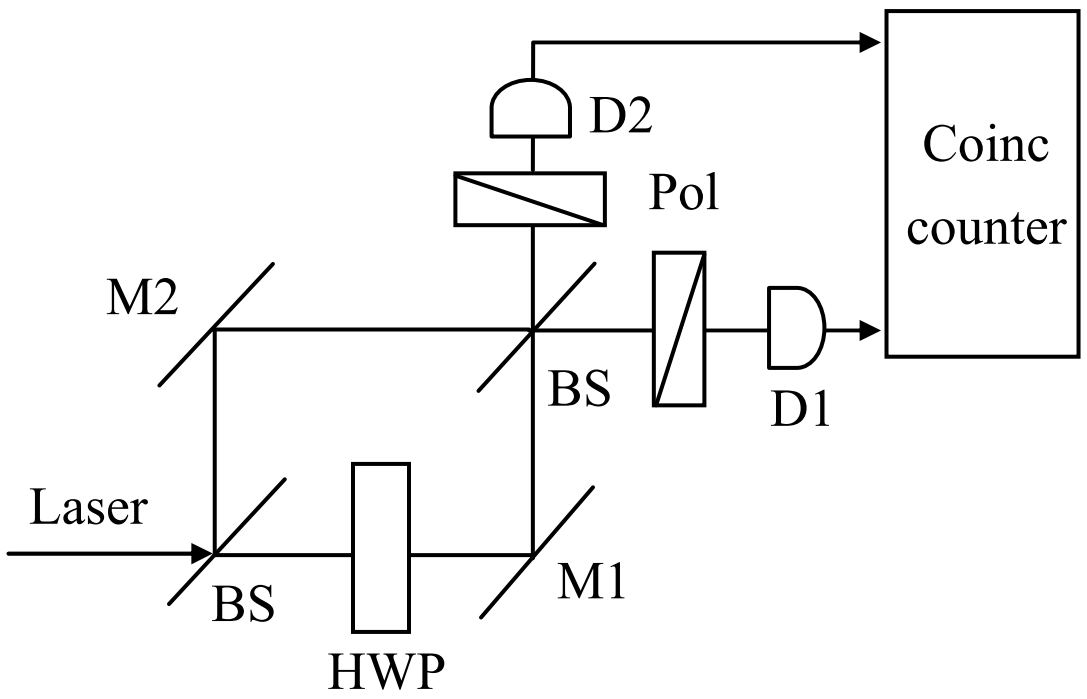

**Fig. 6.** $|\phi^{\pm}\rangle$ state obtained by the overlap of multi-photon wave packets.

In fact, there is a simplest way to generate polarization entangled Bell state. We have supposed that the polarization hidden variables of a bunch of photons within the coherence length are correlated, so if we split a beam of circularly polarized light into two and detect them within the coherence time of laser, Bell state will be obtained. The experimental setup is shown in Fig. 7. A 50/50 beam splitter is used to split the circularly polarized laser. The two beams of light are then analyzed by polarizers. Note that if a half-silvered mirror is used as beam splitter, then different placements of the mirror will result in different Bell states. If reflection occurs at the front of the mirror, a phase shift of $\pi$ accompanies the reflected beam, the rotation directions of the two beams of light are opposite, then the experiment of Fig. 7 will generate $|\phi^-\rangle$. While $|\phi^+\rangle$ state will be obtained in the case of rear surface reflection, since the medium behind the mirror (air) has a lower refractive index than the medium the light is traveling in (glass). If our above prediction is true, then arbitrary number of correlated photons can be obtained by use of a couple of beam splitters as desired. For comparison, thus far the maximum number of entangled photons is eight [35,36].

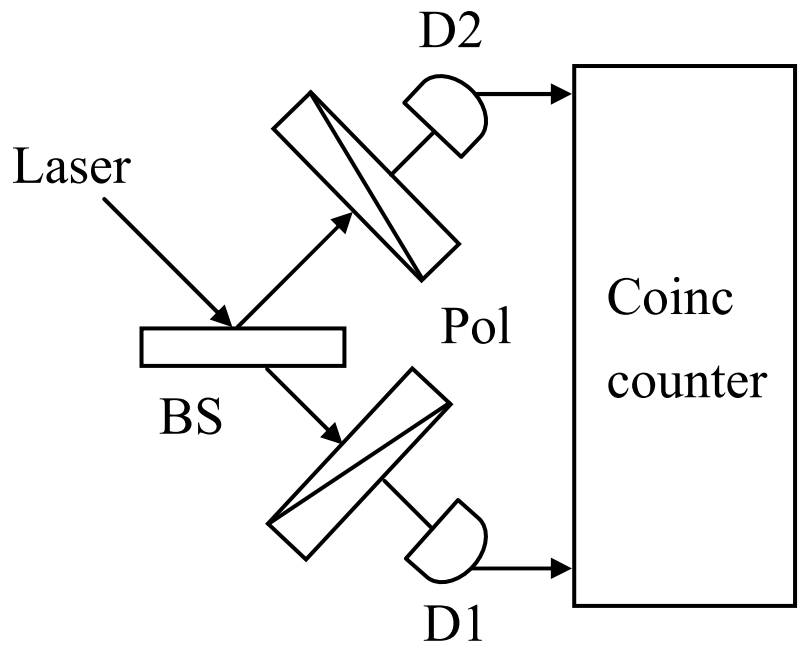

**Fig. 7.** The simplest way to generate polarization entangled Bell state.

Under ideal conditions, all the photons within the coherence length will behave in the same way under the same measuring condition, i.e. they will act as one photon. However, there exists a major difficulty for the experiments of Figs. 6 and 7. Due to the imperfectness of the correlated photons, the experimental setup and the external circumstances, most of the photons will behave in the same way while a few of them may not. This will blur the experimental results. In order to overcome this problem, the single-photon detectors D1 and D2 may be replaced with photoelectric detectors. Note that in this case, our concern is not the specific values of the intensities of the laser but their relative values compared to the threshold value. When all the photons pass through the polarizer, the detection intensity will be that of half of the input laser. So the threshold value may be set at one-fourth of the intensity of the input laser. If both the luminosities in the two channels are greater (or less) than the threshold value, we get a coincident count.

Note that the above experiment differs from that of Hanbury Brown and Twiss (see [37]) in that the former detects the polarization correlation while the latter detects the intensity correlation of input laser. In the meanwhile, it is interesting to compare our experiment with that of Refs [38] and [39], where a polarization beams splitter is used to seperated a beam of monochromatic light (or two orthogonally polarized photons generated by type-II down-conversion), then the relationship between intensity correlation (or coincidence count) and the rotation angle of the half-wave plate placed in front of the polarization beams splitter is recorded. In order to explain the correlated results that cannot be explained in classical theory, hidden polarization or higher-order polarization is introduced [38-40]. We think that they are equivalent to the polarization hidden variable we introduce here. In fact, if two quarter-wave plates are inserted after the polarization beam splitter in the experiments of Refs [38] and [39] to convert the linearly polarized photons into circularly polarized photons, as is the case in Fig. 5, the two experiments will also generate polarization Bell state.

### *4.3. Successive polarization measurements on EPR photon pairs*

If quantum measurement is deterministic, then the experimental result is determined by measuring condition and intrinsic property of a particle, and there are no random disturbances during the measuring process. We may further infer that the collapsed quantum states of a pair of particles with a same quantum state will be the same under the same measuring condition. We now test this assumption. We add another pair of polarizers II and II' in the transmitted channels of Fig. 2, as shown in Fig. 8. The polarizer I has the same orientation as polarizer I', and the orientations of polarizers II and II' are also the same. The source generates circularly polarized $|\phi^+\rangle$ state photon pairs. According to Eq. (9), half of the photon pairs will pass through the first pair of polarizers and reach the second pair of polarizers. When they are analyzed again, their behaviors are still correlated, i.e. if one photon is transmitted, the other will also be transmitted. Thus for the second pair of polarizers, we have $P_{++}=\cos^2\theta$, $P_{--}=\sin^2\theta$, $P_{+-}=P_{-+}=0$, where $\theta$ is the angle between the orientations of the two pairs of polarizers. According to quantum theory, the pair of photons is not in entangled state after the first measurement since their polarizations are definite. In this case, we don't know how to calculate the joint probability in quantum mechanics. But if our predications are correct, there will exist conceptual difficulty for quantum mechanics to explain the total correlation of a pair of particles without entanglement, which can be readily understood in deterministic hidden variable theory. Note that we can also perform the experiment in the reflected channels of ploarizers I and I′, and similar results will be obtained. The joint measurements between transmitted and reflected channels are not needed, since the probabilities will be zero according to Eq. (10). Thus the experiment is a complete measurement.

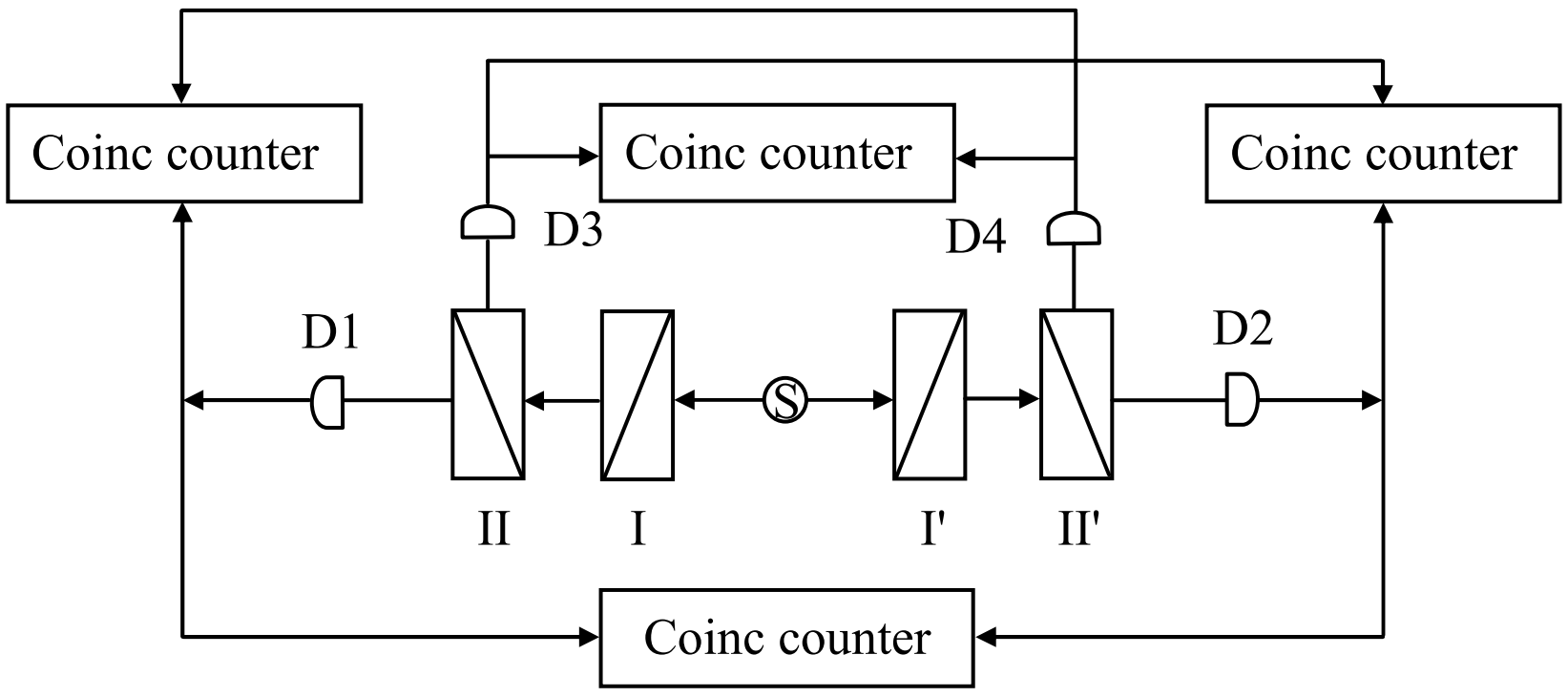

**Fig. 8.** Two successive polarization measurements on EPR photon pairs.

As the collapsed quantum states of a pair of photons after the first measurement are the same, they can be restored into $|\phi^+\rangle$ state by inserting two quarter-wave plates with parallel-oriented optical axes into the optical paths between the two pairs of polarizers.

We now see the coincidence counting results when the orientations of the second pair of polarizers are different. Suppose the orientation of the first pair of polarizers is in the $x$ axis, and the orientations of the second pair of polarizers in the directions of $a$ and $b$, respectively. For simplicity, let $a$, $b$ and $x$ lie in one plane, and $\bar{a}$, $\bar{b}$ are the directions perpendicular to $a$ and $b$, respectively, as shown in Fig. 9.

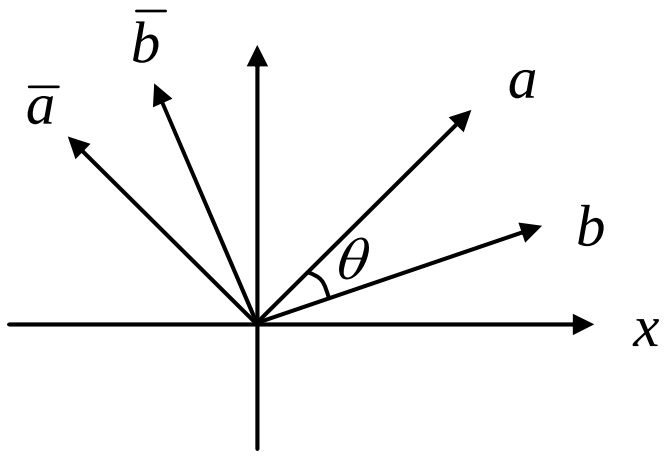

**Fig. 9.** Orientations of two pairs of polarizers.

For the circularly polarized photon pairs, the single probabilities $P_+(a)$ and $P_+(b)$ are equal, we can get the same joint probability $P_{++}(a,b)$ whether by projecting $a$ onto $b$ or by projecting $b$ onto $a$. For a pair of linearly polarized photons, the single probabilities that the two photons pass through the second pair of polarizers respectively are not equal. Then different projective sequences will lead to different results. If we project $a$ onto $b$, we get $P_{++}(a,b)=\cos^2 a\cos^2\theta$, where $\theta=a-b$. If we project $b$ onto $a$, we obtain $P_{++}(a,b)=\cos^2 b\cos^2\theta$. As joint probability cannot be larger than single probabilities, and the latter may not satisfy this requirement, we choose $P_{++}(a,b)=\cos^2 a\cos^2\theta$ for the moment.

We now consider the expression of $P_{+-}(a,b)$. According to the rule of projecting from one channel with a smaller probability onto the other with a larger probability, we obtain $P_{+-}(a,b)=\cos^2 a\sin^2\theta$ for $\cos^2 a\le\sin^2 b$ and $P_{+-}(a,b)=\sin^2 b\sin^2\theta$ for $\cos^2 a\ge\sin^2 b$. As the requirement of $P_{++}(a,b)+P_{+-}(a,b)=P_{+}(a)=\cos^2 a$ must be satisfied, and considering the smooth joining of probability formula, we take

$$P_{++}(a,b)=\begin{cases}\cos^2 a\cos^2\theta & \cos^2 a\le\sin^2 b\\ \cos^2 a-\sin^2 b\sin^2\theta & \cos^2 a\ge\sin^2 b\end{cases}. \qquad (14)$$

It can be verified that in addition to satisfying the projective relation in the instances of $\theta=0$ and $\theta=\pi/2$, Eq. (14) also meets the expectations of $P_{++}(a,b)=\cos^2 a$ for $b=0$ and $P_{++}(a,b)=0$ for $a=\pi/2$. So it is a reasonable probability formula. With Eq. (14) we can calculate the other three joint probabilities using the relations of $P_{++}(a,b)+P_{+-}(a,b)=\cos^2 a$, $P_{++}(a,b)+P_{-+}(a,b)=\cos^2 b$ and $P_{+-}(a,b)+P_{--}(a,b)=\sin^2 b$.

In fact, there may exist other projective relations for the calculation of joint probability. When $b$ rotates between 0 and $a$, the joint probability $P_{++}(a,b)$ may remain unchanged and is always $\cos^2 a$, i.e. joint probability takes the smaller one of the two single probabilities. This implies that for two dependent events under certain conditions (for example, $a$ and $b$ lie in the same quadrant), if one event with a smaller probability occurs, then another event with a larger probability will occur with certainty. Then the four joint probabilities can be written as

$$\begin{cases}P_{++}(a,b)=\cos^2 a\\ P_{+-}(a,b)=0\\ P_{-+}(a,b)=\cos^2 b-\cos^2 a\\ P_{--}(a,b)=\sin^2 b\end{cases}. \qquad (15)$$

It can be seen that in the instance of $\theta=0$ we get the same result as Eq. (14), i.e. $P_{++}(a,b)=P_{+}(a)=P_{+}(b)=\cos^2 a$. In other cases, we cannot decide whether Eq. (14) or (15) is correct, which can only be tested by experiment. No matter which formula is correct, we believe that for a deterministic measurement theory, the requirement that joint probability equals the single probabilities must be satisfied in the case of $\theta=0$.

If we suppose the polarization direction (the $x$ axis in Fig. 9) of photon pairs distributes uniformly in space and then average over it to get average joint probability, we find that whether the result of Eq. (14) or (15) will not agree with that of quantum mechanics. If the polarization direction of photons distributes in two orthogonal directions, the result also disagrees with that of quantum mechanics. We do not present the detailed calculation process. So linearly polarized photon pairs cannot form a Bell state in the general case.

## 5. Discussion and conclusion

We show that the true implication of the probability distribution of Bell's correlation function is the probability distribution of the joint measurement outcomes, so it may vary with experimental condition. In addition, we show that Bell's locality holds neither for two independent events nor for two dependent events. The results of EPR-type experiments can be explained with the projective relation of the quantum state composed of circularly or linearly polarized photon pair whose hidden variables are maximally correlated or anti-correlated. We also explore the physical meaning of hidden variable and measuring process.

Hidden variable theory does not conflict with the current formalism of quantum mechanics, which can be viewed as holding for the statistic description of the behaviors of a large number of independent particles but not for the deterministic description of the behavior of individual particle or EPR pairs. So far there is no experiment suggested to distinguish between the locality and non-locality assumptions. Our first experiment is aimed at this purpose, which we think can verify whether collapse of the wave packet of EPR pair is true or not. All our expectations for above experiments are based on the assumptions that local hidden variable exists and the behaviors of microscopic particles are also deterministic. But it should be noted that even if all our theoretical expectations are verified by experimental results, we could only abandon the concept of quantum entanglement and Bell's locality assumption. Though the start point of our theory is local hidden variable, the above experiments cannot adequately prove that local hidden variable does exist. Only when the experimental results cannot be explained by the current theory of quantum mechanics can we say that it is incomplete and hidden variable should be introduced. So more experiments and theoretical analyses are needed in order to solve the problem of hidden variable.